\documentclass[final,5p,times,twocolumn]{elsarticle}

\usepackage{latexsym}
\usepackage{graphicx}
\usepackage{color}
\usepackage{amssymb}
\usepackage{subfigure}
\usepackage{array}
\usepackage{hhline}
\usepackage{amsmath}
\usepackage{MnSymbol}
\usepackage{supertabular}

\journal{Combustion and Flame}

\definecolor{orange}{rgb}{1,0.65,0}

\begin{document}

% *****************************************************************
% FRONT MATTER
% *****************************************************************

\begin{frontmatter}

\title{Thermal-diffusive instabilities in unstretched, planar diffusion flames}

\author[KTH,EPFL]{Etienne Robert\corref{cor1}}
\ead{etienne@mech.kth.se}

\author[EPFL]{Peter A. Monkewitz}
\ead{peter.monkewitz@epfl.ch}

\cortext[cor1]{Corresponding author, tel.+46-8-790-7580, fax +46-8-796-9850.}

\address[KTH]{Department  of Mechanics, Kungliga  Tekniska H\"ogskolan
  (KTH), Osquars Backe 18, Stockholm 100 44, Sweden.}
\address[EPFL]{Laboratory  of  Fluid  Mechanics (LMF),  Swiss  Federal
  Institute   of  Technology   (EPFL),  Station   9,   1015  Lausanne,
  Switzerland.}

% ********
% Abstract
% ********

\begin{abstract}
  The recent development of a novel research burner at EPFL has opened
  the way  for experimental investigations  of essentially unstretched
  planar diffusion  flames. In particular,  it has become  feasible to
  experimentally  validate  theoretical  models for  thermal-diffusive
  instabilities  in idealized  one-dimensional  diffusion flames.   In
  this paper, the instabilities  observed close to the lean extinction
  limit are mapped in parameter  space, notably as function of the two
  reactant  Lewis numbers.  Cellular  and pulsating  instabilities are
  found at low  and high Lewis numbers, respectively,  as predicted by
  linear stability analyses.  The  detailed investigation of these two
  types  of instabilities  reveals  the dependence  of  cell size  and
  pulsation frequency on the transport properties of the reactants and
  on flow  conditions.  The experimental  scaling of the cell  size is
  found  in  good  agreement  with linear  stability.  The  comparison
  between experimental  and theoretical pulsation  frequencies, on the
  other  hand, was  hampered  by the  impossibility of  experimentally
  reproducing the parameters of  the stability calculations. Hence, a
  heuristic   correlation  between   pulsation   frequency  and   flow
  parameters,  transport  properties,  in particular  the  Damk\"ohler
  number,  and oscillation  amplitude  has been  developed and  awaits
  theoretical interpretation.
\end{abstract}

\begin{keyword}

Diffusion flame \sep Unstrained \sep thermal-diffusive instabilities

% 47.54.-r Pattern selection; pattern formation
% 82.33.Vx Reactions in flames, combustion, and explosions 

\PACS 47.70.Pq \sep  82.33.Vx


\end{keyword}

\end{frontmatter}

% \linenumbers

%***************
% MAIN TEXT BODY
%***************

\section{Introduction}
\label{sec:Intro}

Thermal-diffusive  instabilities,  henceforth  abbreviated  TDIs,  are
known to occur in both premixed and non premixed combustion. They have
become  increasingly relevant  in  the quest  to  reduce emissions  by
operating  combustors  with  leaner  mixtures  which  make  them  more
susceptible to TDIs.  These instabilities also play a key role in soot
formation \cite{Tait1999,  Bradley2000} and in  dynamic extinction and
re-ignition processes.   TDIs were first noticed  through the peculiar
cellular structures they can induce in premixed flames, which has been
noticed  over a  century  ago on  Bunsen burners  \cite{Smithells1892,
  Smith1928}   and   later  recognized   as   originating  from   TDIs
\cite{Zeldovich1944, Markstein1949}.   The first experimental evidence
of such instabilities in a non-premixed flame came much later with the
work of  Garside \& Jackson  \cite{Garside1951}. For broad  reviews of
the  field  see  \cite{Shivashinsky1983,  Buckmaster1993,  Clavin1994,
  Matalon2007}.

The difference between TDIs in premixed and non-premixed flames is the
strong  coupling  between  reaction  and  diffusion  in  the  premixed
configuration  which  gives   rise  to  aerodynamic  (Darrieus-Landau)
instabilities resulting from thermal expansion \cite{Shivashinsky1983,
  Clavin1994}. In non-premixed or diffusion flames, on the other hand,
it is possible to reduce or practically eliminate aerodynamic effects.
In   a   recent   study   \cite{Matalon2010},  it   was   shown   that
thermal-diffusive  processes are  at  the origin  of instabilities  in
diffusion flames with thermal  expansion having only a minor influence
on their  onset.  This relative  simplicity has spurred  an increasing
number    of    stability     analyses    of    non-premixed    flames
\cite{Matalon2007}.  However,  non-premixed flames without aerodynamic
effects are not just handy to analyze but are relevant to the modeling
of  turbulent partially premixed  combustion.  In  the context  of the
turbulent  flamelet   model  \cite{Peters1988},  where   chemistry  is
considered  fast compared  to  the transport  processes, the  reaction
occurs  in  thin layers.   However,  despite  the asymptotically  thin
reaction zone, it can be shown that the flamelet dynamics is dominated
by diffusion  processes and that  advection is a higher  order effect.
Hence, TDIs  are likely to play  a role in the  extinction dynamics of
flamelets.

\begin{figure}[ht!]
  \begin{center}
    \includegraphics[width=0.35\textwidth]{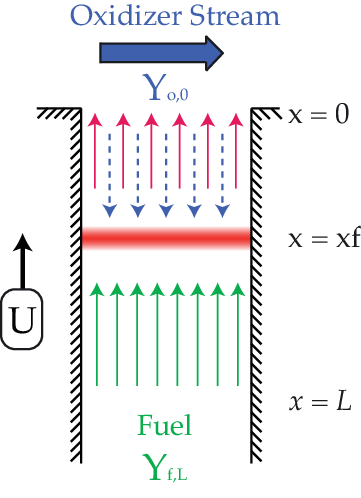}
  \end{center}
  \caption[Conventional  chambered  diffusion  flame]{Schematic  of  a
    chambered diffusion  flame with upward  bulk flow of  velocity $U$
    and  1-D flame  at  $x_f$.   Fuel (green)  is  supplied through  a
    semi-permeable  membrane  at  the  bottom, while  oxidizer  (blue)
    diffuses against the bulk flow of products (red).  Here and in the
    following figure transport by the  bulk flow is indicated by solid
    arrows while  broken arrows  represent diffusion against  the bulk
    flow.}
  \label{fig:1DBurnerSchem}
\end{figure}

The  simplest   possible  non-premixed  flame   is  a  one-dimensional
diffusion flame with or without a uniform bulk flow normal to the flat
unstrained flame.   Such a simple one-dimensional  diffusion flame has
first been introduced as a theoretical construct by Kirkby and Schmitz
\cite{Kirkby1966}.      A    sketch     is     provided    in     Fig.
\ref{fig:1DBurnerSchem}.   The combustion chamber  is a  straight duct
open at one end to a fast  stream of oxidant and supplied at the other
end  with  fuel  through  a semi-permeable  membrane.   All  transport
processes  are  one-dimensional  with  the  oxidant  counter-diffusing
against the flow  of products to the planar reaction  sheet. In such a
simple  configuration,  analytical  solutions  can be  found  for  the
suitably simplified steady  governing equations and numerous stability
analyses  of  this system  have  been  carried out  \cite{Matalon1979,
  Kim1996,  Kim1997, Cheatham2000,  Metzener2006}.   All these  models
predict a planar stable flame  sheet when far from extinction, and the
possibility  of thermal-diffusive  instabilities when  approaching the
lean extinction limit. The type  of instability, if any, is a function
of many parameters, the most important being the Lewis numbers of both
reactants.    The   problem   with   the   idealized   one-dimensional
configuration    studied   theoretically    and   shown    in   Fig.
\ref{fig:1DBurnerSchem}  is   that  it  is   impossible  to  implement
experimentally because  a fast  stream over a  cavity does  not supply
reactant and remove products uniformly over the cross section.

% Figure Burner Configuration
% Combine or add fig. 3.1 and fig. 4.5

Therefore, experimental  studies have until recently  been carried out
in   burners  that  mimic   the  idealized   1-D  burners   of  Fig.
\ref{fig:1DBurnerSchem} only approximately. One popular configuration
is the opposed  jet flame which is flat  but experiences flame stretch
due  to the  elimination  of  products parallel  to  the flame.   This
configuration has been used  to investigate the stability of diffusion
flames \cite{Katta2000} as  well as the effect of  strain on chemistry
\cite{Du1995,  Brown1998}  and  extinction limits  \cite{Pellett1998}.
For a review of various counterflow diffusion flame configurations see
\cite{Tsuji1982}.  Diffusion flames  with particularly low strain have
been  generated  close  to  the  forward stagnation  point  of  porous
cylinders and  hemispherical caps injected  with fuel and placed  in a
slow stream  of oxidant.   With great care,  the strain rate  in these
configurations could  be made as low as  $1.4$ s$^{-1}$ \cite{Han2005}
and  the   thermal-diffusive  instabilities  observed   close  to  the
extinction  limit in  this type  of burner  agreed  qualitatively with
numerical models \cite{Nanduri2005}.  However, in all these burners it
is  conceptually  impossible   to  completely  eliminate  hydrodynamic
effects.

\begin{figure}[t!]
  \includegraphics[width=0.48\textwidth]{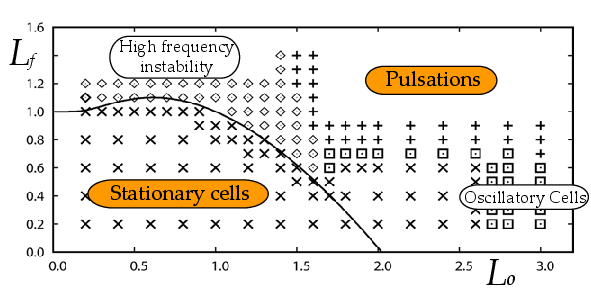}
  \caption[Detailed  map  of   the  instabilities  expected  close  to
  extinction.]{Map of  linear instabilities  close to extinction  as a
    function of fuel and oxidizer Lewis numbers for an initial mixture
    strength $\phi=0.5$. Reproduced from \cite{Metzener2006}.}
  \label{fig:LewisMap}
\end{figure}

Therefore, a  novel research burner  configuration is used  here which
mimics  the idealized  configuration  of Fig.  \ref{fig:1DBurnerSchem}
with truly \emph{unstretched} flame sheets. It is used here in a first
attempt  of  directly  comparing  experimental  observations  and  the
aforementioned stability analyses. In the following, the comparison is
restricted  to  the most  comprehensive  study  of  both cellular  and
pulsating   instabilities    to   date   by    Metzener   \&   Matalon
\cite{Metzener2006}  who  investigated  a  wide range  of  parameters,
including the two independent Lewis  numbers not limited to unity, the
initial  mixture strength  as  well  as the  bulk  flow magnitude  and
direction.   Their  instability  map  is  reproduced  here  as  Fig.
\ref{fig:LewisMap}  to  illustrate  the  rich  behavior  of  even  the
simplest 1-D flame,  with cells at small Lewis  numbers and pulsations
dominating at large Lewis numbers, typically above unity.

% Figure: Stability map from Metzener2006

The  paper   is  structured   as  follows:  first,   the  experimental
methodology is  presented and an  example of stability  and extinction
limits  is shown  as a  function of  the burner  operating parameters.
Then, the type  of instability occurring close to  the lean extinction
limit is identified as a function  of the Lewis numbers for both H$_2$
and CH$_4$  flames.  Finally  the two types  of TDIs observed  in this
experiment,  cellular  flames  and  planar intensity  pulsations,  are
described  in  detail  and   their  characteristics  are  compared  to
theoretical predictions.

\section{Experimental approach}
\label{sec:Exp}

\begin{figure}[t!]
  \begin{center}
    \includegraphics[width=0.48\textwidth]{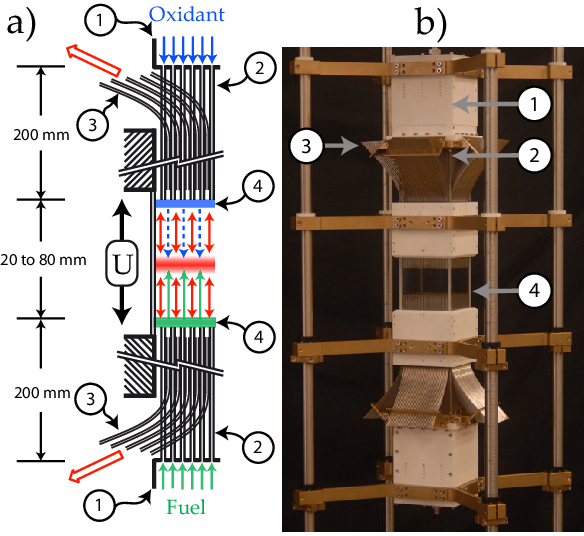}
  \end{center}
  \caption[Schematic and photograph  of symmetric Mark II burner.]{(a)
    Schematic of the  symmetric Mark II burner. Only  the left half is
    shown with bulk flow up and  the same color scheme as in Fig. 1:
    (1),  Reactant injection plenums;  (2), Straight  injection needle
    arrays  between plenums  and combustion  chamber;  (3), Extraction
    needles  bent   outwards  between  injection   needles;  (4)  Thin
    injection layers at the tip  of the needle arrays.  (b) Photograph
    of  the partially  assembled burner  (without exhaust  plenums and
    thermal insulation) with all needle arrays in place.  (1), (2) and
    (3), same as in (a); (4), Quartz-walled burning chamber.}
  \label{fig:MarkII}
\end{figure}

The main  difficulty of creating  an unstrained chambered flame  is to
supply the reactants and remove  the products evenly across the burner
cross section. This problem  has been successfully solved by supplying
reactants  and  removing products  through  closely  spaced arrays  of
hypodermic  needles \cite{LoJacono2005, Robert2008,  Robert2009}.  The
latest symmetric  Mark II  version of the  burner used to  produce the
present  results is  shown  in Fig.  \ref{fig:MarkII},  with a  more
detailed  description available  in \cite{Robert2009}.   The principal
characteristic of  this burner is the supply  of \emph{both} reactants
through  arrays of  hypodermic  needles and  the  possibility for  the
products to escape  on either side of the  combustion chamber allowing
an independent  control of the  bulk flow direction and  magnitude. In
this study, however,  the fuel is always injected  from the bottom and
the products escape only at  the top.  The injection arrays consist of
$31^2=961$ stainless steel tubes ($1.2$  mm O.D., $1.0$ mm I.D.)  on a
$2.5$ mm Cartesian grid. Typical  Reynolds numbers in the supply tubes
and in the  burning chamber (based on chamber width)  are on the order
of $10$ and $50$, respectively.

\begin{figure}[ht!]
 \begin{center}
  \includegraphics[width=0.35\textwidth]{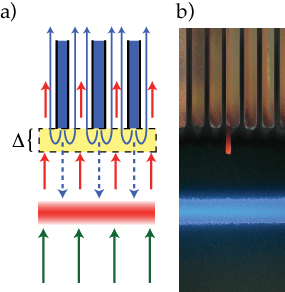}
  \end{center}
  \caption[Detailed   representation  of   the   injection  layer.]{a)
    Schematic  of the  upper injection  layer (shaded  area, thickness
    $\Delta$) immediately below the oxidizer injection array. b) Photo
    of  the mass  spectrometer capillary  inserted through  an exhaust
    tube to probe the mixture composition below the injection layer.}
  \label{fig:InjLayer}
\end{figure}

To ensure an even pressure drop  in the exhaust path across the burner
cross-section,  the combustion  products are  removed  through another
needle array located within the injection array and bent outwards. The
discrete nature of  the reactant supply implies that  thin layers with
localized  three-dimensional flow  exist next  to the  tip of  the two
needle  arrays.   They are  referred  to  as \emph{injection  layers}.
Between these two  thin layers the species transport  is very close to
one-dimensional  and the flame  experiences minimal  residual stretch.
The thickness of  these layers is obviously strongly  dependent on the
burner operating  conditions, but for the flames  investigated here it
has  been  measured  to  be   of  the  order  of  the  needle  spacing
\cite{Robert2009}. Also,  a significant  portion of oxidizer  is swept
directly into the exhaust instead of diffusing against the products as
shown  in  Fig.  \ref{fig:InjLayer}.   In  order  to  determine  the
effective  boundary  conditions outside  the  injection  layer of  the
counter-diffusing  species,  here  the  oxidizer,  and  the  transport
properties at the flame,  a carefully calibrated mass spectrometer was
used to  analyze samples collected  throughout the burning  chamber in
real-time.  The accuracy  of this technique has been  determined to be
between  5\% and  7\% for  most of  the mixtures  used in  this burner
\cite{Robert2010}.

The   phenomena  responsible  for   perturbing  the   idealized  flame
configuration of Fig. \ref{fig:1DBurnerSchem} have been identified,
minimized and quantified and  are described in \cite{Robert2009}.  The
most  significant  of  these  effects  results  from  the  temperature
gradient  induced by  the presence  of  the walls  around the  burning
chamber.  Using radiation  corrected and ceramic coated thermocouples,
the flame temperature close to  the chamber walls has been measured to
be as much as 350K below  the temperature in the central region.  This
lower flame  temperature at the  periphery of the burner  implies that
the oxidant supply needles located near the periphery of the array are
also colder. As a result, the flow of oxidant in the first few rows of
supply needles  is less viscous and  enters the chamber  with a higher
velocity.   The net  effect, measured  with mass  spectrometry,  is to
decrease the mixture strength at  the periphery of the burning chamber
and hence curve the flame edges down, away from the oxidizer injection
array. However, in  the central $2/3$ of the burner  where the TDI are
investigated  the flame  front deviated  by  less than  $0.5$ mm  from
perfect  flatness, with  the  flame temperature  and mixture  strength
deviating by less than $10\%$ from their average values.  By combining
velocity  profiles collected  by LDA  with the  shape of  the reaction
front, the residual  stretch experienced by the flame  was found to be
about $0.05$  s$^{-1}$ \cite{Robert2009} in  the central $2/3$  of the
flames and $0.15$ s$^{-1}$ in the curved edge regions.

Another difficulty is that in  theoretical analyses such as the one by
Metzener  and Matalon \cite{Metzener2006},  the Damk\"ohler  number is
conveniently  controlled  by  changing  the bulk  velocity  $U$  while
keeping  all other  parameters  constant.  Unfortunately  this is  not
possible experimentally  in the  present configuration because  of the
strong coupling  between $U$ and the  flow in the  injection layer, in
particular the fraction of oxidant entrained directly into the exhaust
(see   Fig.   \ref{fig:InjLayer}),   which  defines   the   boundary
conditions.   As  a result,  the  range  of  velocities leading  to  a
configuration   similar    to   the   idealized    model   of   Fig.
\ref{fig:1DBurnerSchem} is very narrow: If the bulk flow is too fast,
the flame is  pushed into the downstream injection layer  and if it is
too low  the flame sheet  is so far  from the upper  oxidant injection
array that buoyancy driven instabilities arise \cite{Robert2008}.  For
the  present burner  design, the  only way  to significantly  vary the
Damk\"ohler  number is  through  the flame  temperature  which can  be
changed either through the dilution  (its nature or its amount) or the
mixture strength.   Changing the burning  temperature through dilution
while maintaining constant mixture strength is very tricky because the
nature of the  inert has a strong influence on  the injection layer of
the counter-diffusing oxidant.   Therefore, in the present experiments
the  Damk\"ohler number  was lowered  simply by  reducing  the mixture
strength via the fuel volume fraction.

In  the  limit  of   infinitely  fast  and  complete  combustion  (the
Burke-Schumann  limit, $Da  \to \infty$)  a one  dimensional diffusion
flame  sheet  is  unconditionally  stable.  Therefore,  the  following
investigations  of instabilities  were all  started from  stable, high
Damk\"ohler  number  baseline  flames  which  have  been  successfully
described by a  simplified one-dimensional model in \cite{Robert2009},
demonstrating that  the actual burner represents  a good approximation
of the idealized configuration of Fig. \ref{fig:1DBurnerSchem}.  By
decreasing $Da$ from the baseline value the stability boundary of TDIs
was  identified  first.   A   further  decrease  of  $Da$  allowed  to
characterize the instabilities until extinction was reached.

\section{Stability and extinction limits}
\label{sec:Map}

\begin{figure}[ht!]
  \begin{center}
    \includegraphics[width=0.48\textwidth]{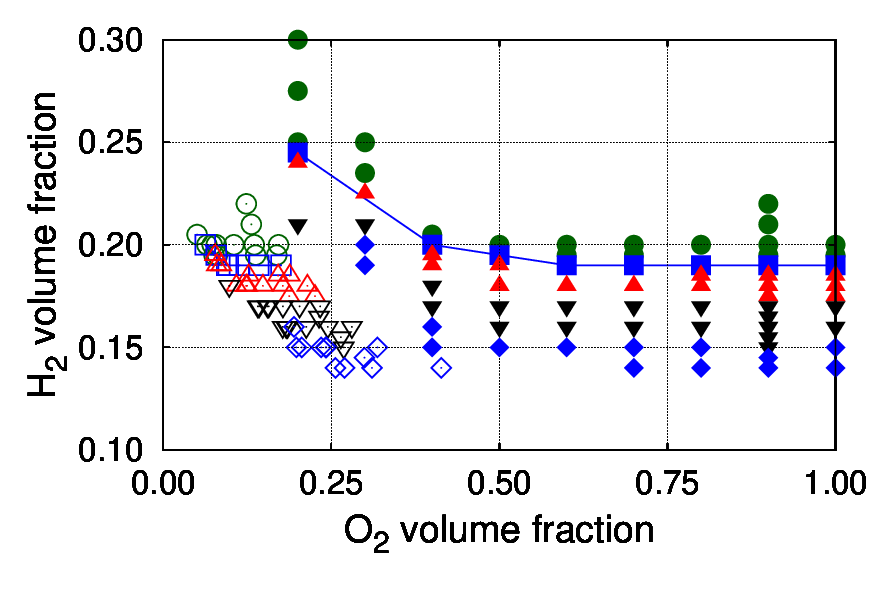}
    \caption[Stability   limits  inferred   from  the   supplied  flow
    rates.]{Flame stability  limits as  a function of  reactant volume
      fractions at a bulk velocity  of $U=19.4$ mm/s, . Solid symbols:
      stability limits versus supplied reactant volume fractions; Open
      symbols: limits  versus reactant volume  fractions obtained from
      mass   spectrometry   outside   the   injection   layer.    Key:
      (\textcolor[rgb]{0,0.4,0}{$\circ$})    Stable   planar   flames;
      (\textcolor{blue}{$\Box$})  Onset of  first holes  in  the flame
      sheet;   (\textcolor{red}{$\bigtriangleup$})   Partly   cellular
      flame;  (\textcolor{black}{$\bigtriangledown$})  Fully  cellular
      flame;   (\textcolor{blue}{$\Diamond$})  Fully   cellular  flame
      collapsed on injection array.}
    \label{fig:StabLimits}
  \end{center}
\end{figure}

% Stability and extinction limits - For H2/O2 flames in CO2
% Discussion on Stability and Extinction limits
In  Fig. \ref{fig:StabLimits},  an  example of  stability limits  is
shown for  a CO$_2$-diluted H$_2$-O$_2$  flame.  To map  the different
flame regimes  as a function  of mixture composition at  constant bulk
velocity,  the following procedure  was used:  Starting from  a stable
flame,  the  fuel  mass   fraction  was  progressively  reduced  while
monitoring flame shape and position  with two cameras. It was observed
that the  flame area  becomes cellular only  gradually as  the mixture
composition is  changed.  This was  expected from the  measurements of
counter-diffusing species distribution  above the flame, as mentioned
in the  previous section.  The different stages  of the  transition to
full    cellularity    are     therefore    documented    in    Fig.
\ref{fig:StabLimits}.

% Figure stability limits, by volume, from flow rates and MS

The  concept   of  extinction  is   also  somewhat  blurred   in  this
configuration since  the flame does not stop  burning completely until
very low fuel volume fractions of the order of $8$ to $10$\% by volume
are reached.  What happens is that, as the flame approaches extinction
and fuel leakage through the  reaction zone increases, the flame front
is displaced  in the direction of  the bulk flow  towards the oxidizer
injection layer.  At some point  the flame then \emph{jumps} into this
highly oxygenated  injection layer where it continues  to burn despite
further  reduction  in  mixture  strength.   Here  the  flame  changes
character as  it is being stabilized  at the tips of  the needles. The
criterion for flame extinction adopted  here is therefore one of flame
position  rather  than one  of  flame  aspect\footnote{Using the  same
  burner  configuration as  here,  \mbox{Lo Jacono}\cite{LoJacono2005}
  defined  in his  thesis the  extinction as  a state  when  the flame
  occupied  only  0.2\%  of   the  burner  cross  section.   From  the
  measurements presented here, this  limit occurs long after the flame
  jumps to  the tips  of the injection  needles, resulting in  a large
  underestimate  of  the   extinction  limit}:  Actual  extinction  is
replaced by the conditions at which  the flame reaches the edge of the
highly  oxygenated   injection  layer   which  is  also   the  virtual
origin\footnote{The virtual origin refers to a position at the edge of
  the  injection  layer  below  which  the burner  can  be  considered
  one-dimensional and  where the boundary conditions for  the 1-D part
  are measured,  see \cite{Robert2008} for details.}   at which $\phi$
is  defined.   In  the  idealized  one-dimensional  burner  of  Fig.
\ref{fig:1DBurnerSchem} this definition corresponds to the flame being
blown out into the top stream.

The  importance  of  defining   reactant  properties  outside  of  the
injection layer(s) is illustrated  by the large difference between the
solid  and  open  symbols  in Fig.  \ref{fig:StabLimits}  where  the
effective volume fractions have  been measured by mass spectrometry at
the edge  of the  oxidizer injection layer  or virtual  origin located
$2.75$  mm below  the injection  array.   The two  datasets of  Fig.
\ref{fig:StabLimits}  demonstrates that  the effective  oxidant volume
fraction at  the virtual  origin is  about $1/3$ of  the value  in the
supply  manifold, i.e.  that $2/3$  of the  supplied oxidant  is swept
directly into the exhaust.

\subsection{Mapping of instabilities}
\label{sec:InstMap}

The Lewis numbers  of both reactants are key  parameters for the onset
of thermal-diffusive  instabilities as they  relate the rate  at which
reactants are supplied  to the heat evacuated from  the reaction zone.
To validate theoretical predictions of the nature of TDIs in diffusion
flames  close to  extinction, one  has  to therefore  explore a  large
enough range of both fuel and oxidant Lewis numbers Le$_f$ and Le$_o$.

Hydrogen-oxygen diffusion flames diluted  in various amounts of CO$_2$
yield only relatively small  Lewis number variations for both species.
For instance,  measurements made in the  Mark I version  of the burner
\cite{Robert2007}  with  such mixtures  only  allowed  Le$_f$ to  vary
between $0.23$  and $0.27$.  In order  to cover wider  ranges of Lewis
numbers, a  mixture of two inerts,  CO$_2$ and He  with very different
transport  properties is  used here  as  dilutant. As  the lighter  He
progressively  replaces  CO$_2$ in  the  dilution  mixture, the  Lewis
numbers  of both  reactants increases  significantly.  This  allows to
investigate a region  of parameter space where both  Lewis numbers are
about  unity or larger  and planar  intensity pulsation  are expected.
The Lewis numbers  of the reactants were calculated  using the Cantera
software  package   \cite{GoodwinXXXX}  which  provides   a  transport
properties model compatible with the multicomponent model described by
Kee  et  al.  \cite{Kee1986}.  As  the  helium  replaces  the  CO$_2$,
H$_2$-O$_2$ flames become  less luminous in the visible  light, to the
point of making observations at high Lewis numbers impossible with the
camera available to  us.  For this reason, a  heavier hydrocarbon fuel
(methane  CH$_4$),  was  used  to  explore  the  parameter  region  of
pulsation  instability.  This  ensured  that CO2  was  present in  the
He-diluted  flames to  enhance  visible light  emission. Although  the
chemical kinetics of  H$_2$ and CH$_4$ are known  to be very different,
the issue is not believed to  be significant here since the two fuels
are not compared and chemical kinetics  do not play a critical role in
the thermal-diffusive instabilities presented here.

\begin{figure}[ht!]
  \begin{center}
    \subfigure[]
    {
      \label{fig:InstMap:a}
      \includegraphics[width=0.35\textwidth]{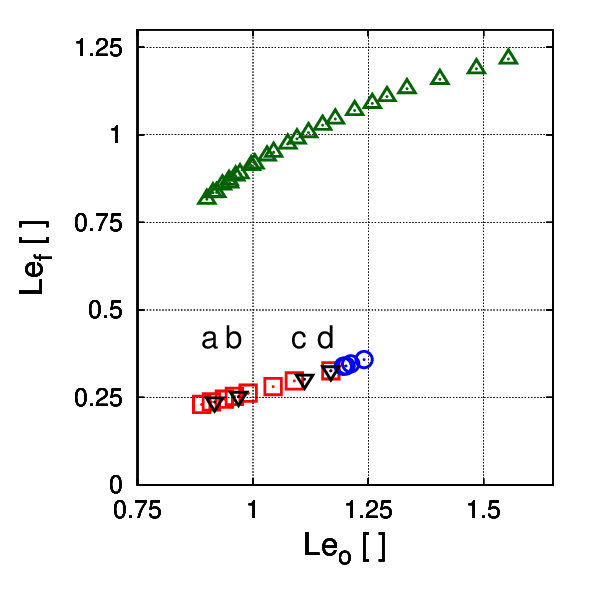}
    }
    \subfigure[]
    {
      \label{fig:InstMap:b}
      \includegraphics[width=0.35\textwidth]{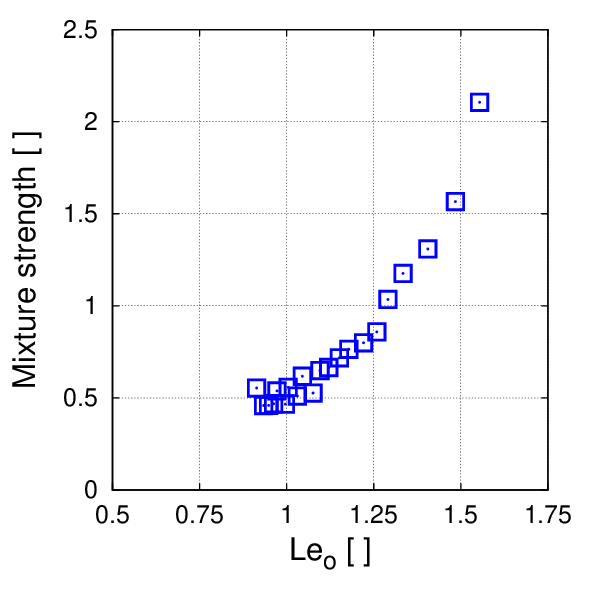}
    }
  \end{center}
  \caption[a)  Mapping  of  the  instabilities  in  the  Lewis  number
  parameter   space.]{  a)   Experimental  instability   map   in  the
    Le$_O$-Le$_f$   parameter  space.    With  \textcolor{red}{$\Box$}
    cellular  hydrogen   flames,  \textcolor{blue}{$\odot$}  pulsating
    hydrogen    flames    and    \textcolor[rgb]{0,0.4,0}{$\triangle$}
    pulsating methane flames. The average mixture strength is 0.45 for
    cellular flames and  the symbols $\bigtriangledown$ represents the
    conditions     for    the     flames    presented     in    Fig.
    \ref{fig:CellScalingDth}.   b) Mixture  strength variation  due to
    inert composition for the pulsating methane flames versus Le$_O$.}
  \label{fig:InstMap}
\end{figure}

The  type of  instability observed  close to  extinction in  the Lewis
number  region covered  by both  H$_2$-O$_2$ and  CH$_4$-O$_2$ diluted
with   a   mixture   of   CO$_2$   and   He   is   shown   in   Fig.
\ref{fig:InstMap:a}.   With  hydrogen diluted  in  CO$_2$ alone,  only
cellular  flames are observed.   When the  CO$_2$ was  almost entirely
replaced by He, the instability  mode of the H$_2$ flames changed from
cellular to oscillatory, but the  very low light emission allowed only
visual observations.  The methane flames, on the other hand, were seen
to  oscillate over  the whole  range of  He$_2$-CO$_2$  mixtures used.

% Comparison with theoretical models  - Mixture strength variations in
% the Lewis number parameter space
The  results  of Fig.  \ref{fig:InstMap:a}  and  the predictions  of
theoretical   models   such  as   those   of   Metzener  and   Matalon
\cite{Metzener2006} are  in relatively  good agreement which  could be
termed \emph{semi-quantitative}.  The discrepancies between the two is
due  to the  impossibility of  experimentally reproducing  exactly the
conditions  used  to  numerically  solve the  theoretical  models,  in
particular  the constant mixture  strength.  The  change in  the inert
composition used  to achieve  the wide range  of Lewis  numbers caused
changes  in  the injection  layer  and  hence  modified the  effective
mixture  strength. Moreover,  since  the flames  were destabilized  by
lowering  the mixture  strength, the  latter could  not  be controlled
independently.    The   value   of   $\phi=0.45$  given   for   Fig.
\ref{fig:InstMap:a} represents  an average value for  all the hydrogen
flames; the  mixture strength  actually varied between  $\phi=0.4$ and
$\phi=0.5$.  This  effect was even  more pronounced for  the pulsating
methane flames as shown in Fig. \ref{fig:InstMap:b}.

% Fig 7.5 : Instability map, Lewis number space

\section{Characterization of observed instabilities}
\label{sec:Instab}

\subsection{Cellular flames}
\label{sec:Cells}

\begin{figure*}[ht!]
  \begin{center}
    \includegraphics[width=0.8\textwidth]{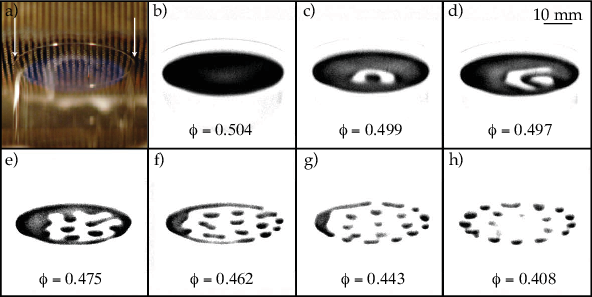}
  \end{center}
  \caption[Photographs of cellular flames.]{a) View of the burner with
    inner cylinder of  48 mm I.D. (edges shown  by white arrows). Same
    field   of   view   as    in   images   b)-h)   and   in   figures
    \ref{fig:CellScalingU}    and    \ref{fig:CellScalingDth}.   b)-h)
    Photographs of  cellular flames with  different mixture strengths,
    from onset  to full cellularity.  H$_2$-flame with  pure CO$_2$ as
    inert  (corresponding to Le$_f$=0.24  and Le$_o$=0.92,  see figure
    \ref{fig:InstMap:a}) and bulk velocity  of $19.4$ mm/s. A video of
    the small  cell motion observed  is available in  the supplemental
    material (video \#1),  corresponding to a flame very  close to the
    one shown in frame h). All flame images inverted for clarity.}
  \label{fig:CellularFlames}
\end{figure*}

An example of  the typical transitions from a stable  flame sheet to a
fully cellular flame as  the mixture strength is progressively reduced
is shown in Fig. \ref{fig:CellularFlames}. The photographs represent
the flame  within a quartz cylinder  of $48$ mm  inner diameter placed
inside  the burner  (visible  in  frame a)  to  isolate the  virtually
one-dimensional  central region of  the chamber  from the  edges where
horizontal  concentration  gradients  remain  \cite{Robert2008}.   The
flame in  the outer part  of the burner  between the cylinder  and the
outer windows  is leaner  and goes through  the same  transitions, but
with strong  cell motion, and generally extinguishes  before the flame
inside the cylinder becomes fully unstable.

The  detailed   characterization  of  this   burner  configuration  in
\cite{Robert2008}  showed  that  the  effective mixture  strength  is
slightly  higher in  the central  portion of  the  burner.  Therefore,
cells are expected to form first  at the leaner periphery of the flame
and then propagate inward as  the mixture strength is reduced. As seen
in  Fig.  \ref{fig:CellularFlames}  the  opposite  is  observed.   A
possible explanation is the increased residual stretch in the velocity
boundary layer  on the quartz cylinder  wall which is known  to have a
strong  stabilizing influence  on thermal-diffusive  instabilities and
may  more than compensate  the destabilizing  effect of  lower mixture
strength.

% Fig. 7.6 : Transition to a fully cellular flame with phi variation
%********(VIDEO OF CELLULAR FLAMES IN SUPPLEMENTARY MATERIAL)*********

\subsection{Cell size scaling}
\label{sec:CellScaling}

\begin{figure*}[ht!]
  \begin{center}
    \includegraphics[width=0.8\textwidth]{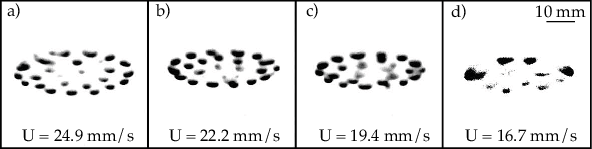}
  \end{center}
  \caption[Cell size scaling  with bulk velocity.]{Photographs showing
    variation   of  cell  size   with  bulk   velocity.  Fuel-advected
    configuration with pure CO$_2$  as inert. Average mixture strength
    $\phi=0.62$.    Le$_f$=0.24    and    Le$_o$=0.92,   see    Fig.
    \ref{fig:InstMap:a}.}
  \label{fig:CellScalingU}
\end{figure*}

% Theoretically predicted cell size scaling
From  theoretical models,  the  size of  the  transverse cell  spacing
$\lambda_c$ in  the cellular structure  is expected to  scale linearly
with  the  diffusion  length $l_D=D_{th}/U$  \cite{Cheatham2000},  the
ratio of the  thermal diffusivity $D_{th}$ to the  bulk flow velocity.
The results of the linear  stability analysis show that the wavelength
with the  maximum growth rate  at marginal stability is  indicative of
the resulting cell size  \cite{Metzener2006}.  Hence, the cell size is
given by

\begin{equation}
\lambda_c \sim \frac{2 \pi l_D}{\sigma^{\ast}}
\label{eq:CellSize}
\end{equation}

where  $\sigma^{\ast}$ is  the preferred  wavenumber at  the  onset of
instability   made  non-dimensional  with   $l_D$.   It   is  obtained
numerically  from  the dispersion  relation  as  a  function of  $Da$,
$Le_o$, $Le_f$, $U$  and $\phi$.  For the lean  flames where cells are
observed in  this burner ($\phi  \approx 0.5$) $\sigma^\ast$  has been
shown \cite{Metzener2006}  to increase only slightly  with $Le_o$ when
far from  a transition to  another instability modes.   However, since
the dependence of $\sigma^\ast$ on  the other parameters which vary in
the present experiment has  to date not been investigated, $\lambda_c$
will in  the following only  be correlated to $l_D$.   To investigate
the  scaling  of  the  cell  size, the  two  quantities  defining  the
diffusion length  $l_D$ were varied independently.   The bulk velocity
is easily changed,  and the effect on the  cellular flame structure is
seen in Fig. \ref{fig:CellScalingU}.   The cell size is clearly seen
to  increase as the  bulk velocity  is decreased,  as expected.  It is
quantified by counting the number of cells around the periphery of the
inner  cylinder  shown   in  Fig.  \ref{fig:CellularFlames}  a)  and
dividing the circumference of the  circle through the centers of these
peripheral cells by the number  of cells. Note that for the conditions
of Fig. \ref{fig:CellScalingU}, the cell pattern starts to rotate at
about one  revolution per second when  the bulk flow  is lowered below
about $18$  mm/s. The bulk flow  velocities given in  this section are
based on  the supplied  (cold) gas flow  rates with  the corresponding
values at the flame typically  about 4 times higher depending on flame
temperature.

% Fig. 7.9 : Cell size scaling with U

\begin{figure}[ht!]
  \begin{center}
    \includegraphics[width=0.45\textwidth]{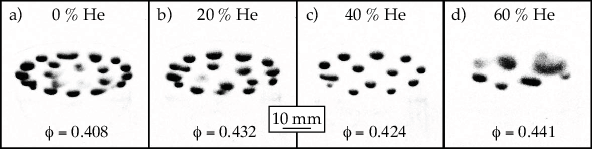}
  \end{center}
  \caption[Cell  size  scaling  with thermal  diffusivity.]{Images  of
    cells  for different  $D_{th}$. H$_2$-advected  configuration with
    $U=19.4$ mm/s. The Lewis numbers for the four frames are indicated
    in Fig. \ref{fig:InstMap:a}.}
  \label{fig:CellScalingDth}
\end{figure}

The  second  parameter  defining  the diffusion  length,  the  thermal
diffusivity, has  been modified by using different  mixtures of CO$_2$
and He$_2$  as diluting  inert. The resulting  change in  the cellular
pattern as  the inert composition  is progressively changed  from pure
CO$_2$ to pure helium is shown in Fig. \ref{fig:CellScalingDth}.  As
expected, the increase of  thermal diffusivity results in an increased
cell size. It is noted here that  the larger size cells in frame d) of
Fig. \ref{fig:CellScalingDth} have small amplitude and show unsteady
size oscillations.  Looking  at the location of this  last frame d) in
the Le$_O$-Le$_f$ plane of Fig. \ref{fig:InstMap:a} reveals that the
somewhat  unsteady  large  cells   are  produced  very  close  to  the
transition to intensity pulsations.  In this region of parameter space
instability mode  competition is expected  \cite{Metzener2006} and has
indeed been  observed sporadically in the present  experiment.  In the
transition  region  between  cells   and  pulsations  the  flame  also
occasionally took  the form of a  rotating planar spiral  which can be
viewed as a  single cell rotating in the present  setup with about one
revolution per second.

% Fig. 7.10 : Cell size scaling with Dth and Lewis

\begin{figure}
  \begin{center}
    \includegraphics[width=0.48\textwidth]{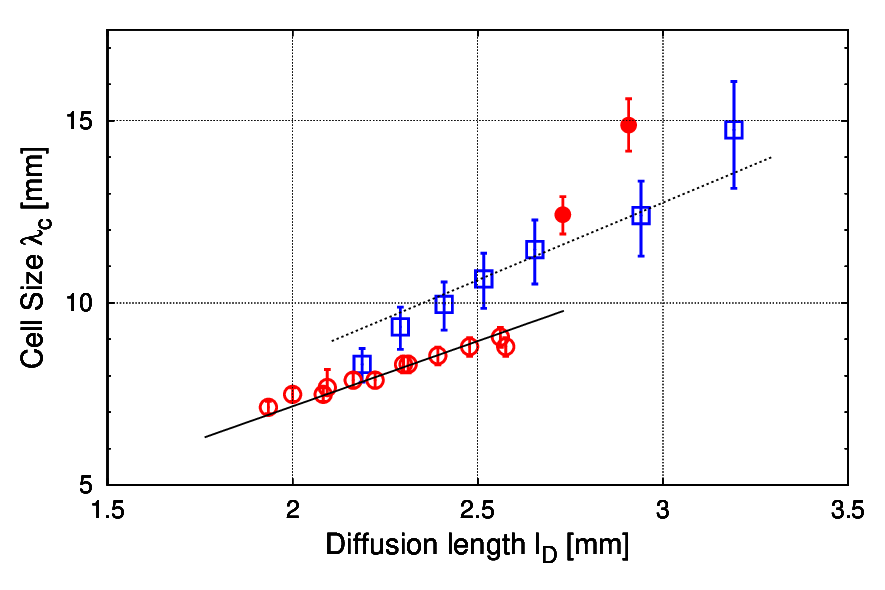}
  \end{center}
  \caption[Scaling of  the cell size.]{Measured  cell size $\lambda_c$
    as a function diffusion  length $l_D$, determined with the thermal
    diffusivity     and     bulk     velocity    at     the     flame.
    (\textcolor{blue}{$\Box$},    --    --    --   corresponding    to
    $\sigma^\ast=1.75$), variable  inert composition at  constant bulk
    velocity  (average  $\phi=0.45$); (\textcolor{red}{$\circ$},  ---
    corresponding  to  $\sigma^\ast=1.5$, \textcolor{red}{$\bullet$}),
    variable  bulk  velocity  with   pure  CO$_2$  as  inert  (average
    $\phi=0.62$),  the solid  points  representing oscillating  cells.
    The error bars represent only the \emph{quantization error} due to
    integer number of cells forming  around the periphery of the inner
    quartz cylinder.}
\label{fig:CellScaling}
\end{figure}

The  cell sizes $\lambda_c$  obtained from  both the  experiments with
variable   $U$   and   variable   $D_{th}$   are   shown   in   Fig.
\ref{fig:CellScaling} as  function of the  diffusion length calculated
with  the   thermal  diffusivity  and  bulk  velocity   at  the  flame
temperature. As  was demonstrated by Jomaas  et al. \cite{Jomaas2007},
the use of locally evaluated values for the transport properties is of
critical importance in such circumstances. The two datasets follow the
linear relationship  between diffusion  length and cell  size expected
from   equation  \ref{eq:CellSize}.  The   different  slopes   can  be
attributed to slightly different average mixture strengths between the
two datasets, on  the order of $\phi=0.45$ for  constant bulk velocity
and  $\phi=0.62$ for  constant inert  composition (and  variable $U$).
The  resulting values  for $\sigma^\ast$  are about  $1.75$  and $1.5$
respectively,  which is  consistent with  the result  of  Metzener and
Matalon \cite{Metzener2006}  that $\sigma^\ast$ is close  to unity for
their lean flames ($\phi=0.5$, but  with $Le$ different from ours) and
increases with $\phi$.

The   major  trends   regarding   the  Lewis   number  dependence   of
$\sigma^\ast$  predicted by this  theoretical investigation  were also
observed  experimentally. First,  when far  from a  transition  in the
instability mode, $\sigma^\ast$ is  not expected to vary significantly
with  $Le_o$,  which  is  confirmed  by  the  quality  of  the  linear
correlations presented in  Fig. \ref{fig:CellScaling}.  Second, when
the instability mode changes from stationary cells to pulsating cells,
as    observed   for    the   variable    $U$   dataset    of   Fig.
\ref{fig:CellScaling},   $\sigma^\ast$   drops  significantly.    This
transition and the associated  decrease of $\sigma^\ast$ were observed
in  the theoretical  results  for lean  flames  ($\phi=0.5$) found  in
\cite{Metzener2006}.

One has to keep in mind however that the theoretical data available in
this reference  was provided for  $Le_f=0.7$ while in  our experiments
$Le_f$ varied between $0.23$ and $0.40$, preventing for the moment the
quantitative validation of the model. Qualitative comparison is on the
other  hand  rather  good,  with the  authors  of  \cite{Metzener2006}
concluding that the ratio  $\lambda_c/l_D$ is expected to vary between
about $3$ and  $12$, depending on the mixture  strength and flow rate.
In  our experiments, this  ratio was  measured in  the $3.4$  to $4.6$
range,  our experimental  facility  only allowing  the observation  of
rather lean cellular flames.   Additionally, the number of data points
was   insufficient   to  correlate   the   deviations  from   equation
\ref{eq:CellSize} with $Le_o$, $\phi$  or $U$, for instance (note that
the Damk\"ohler number is difficult  to include in this list since the
fragmentation of  the flame front precludes  a reliable quantification
of mixture composition and flame temperature).

% Fig. 7.11 : Cell size scaling

\subsection{Planar intensity pulsations}
\label{sec:Pulsations}

\begin{figure*}
  \begin{center}
    \includegraphics[width=0.8\textwidth]{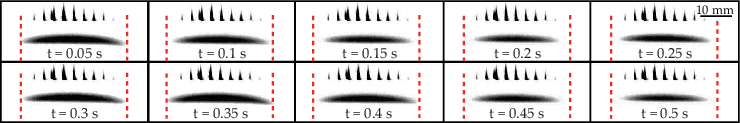}
  \end{center}
  \caption[Pulsation in a methane flame with the inner quartz cylinder
  present.]{Pulsation  of  a methane  flame  within  the inner  quartz
    cylinder.  The frame  width is equal to the  inner diameter of the
    quartz   cylinder  ($48$mm).    Bulk   velocity  $U=19.42$   mm/s,
    $Le_f=1.10$ and  $Le_o=1.20$.  The vertical dashed  lines mark the
    maximum  flame  diameter during  the  cycle  and  the tip  of  the
    injection  array  is  visible  at  the top  of  the  frames.   The
    corresponding  video  is available  in  the supplemental  material
    (video \#2), at one half the original speed.}
  \label{fig:CH4Puls_Cyl}
\end{figure*}

% 8 bit intensity signal
In   this  section  flames   with  high   Lewis  numbers   ($>1$)  are
investigated, in which planar  intensity pulsations are observed close
to extinction.  The  flame dynamics are captured from  the side with a
high-speed video camera (Photron  Fastcam APX) operated between 50 and
250 fps (frames per second).  Since no filtering optics were used, the
signal acquired is the sum of  all the visible light emission from the
flame,  integrated over  the burner  cross-section.  An  example  of a
pulsating methane flame is shown in Fig. \ref{fig:CH4Puls_Cyl}. This
sequence shows  the flame  inside the quartz  cylinder placed  in the
combustor chamber as  shown in Fig. \ref{fig:CellularFlames}a).  The
presence of  this cylinder which  encloses the most uniform  region of
the  burner, results  in the  whole flame  sheet  reaching instability
simultaneously  as  the  mixture  strength  is  decreased.   Following
transition to  instability, the pulsations quickly  grow in amplitude,
often resulting in extinction within a few seconds.

% Fig. 7.15 : Pulsation in cylinder
% Video 14.5.3.1: U=19.42 mm/s, Lef=1.10, Leo=1.20, freq=
%**********(MAKE VIDEO AVAILABLE IN SUPPLEMENTARY MATERIAL)**********

\begin{figure}
  \begin{center}
    \includegraphics[width=0.45\textwidth]{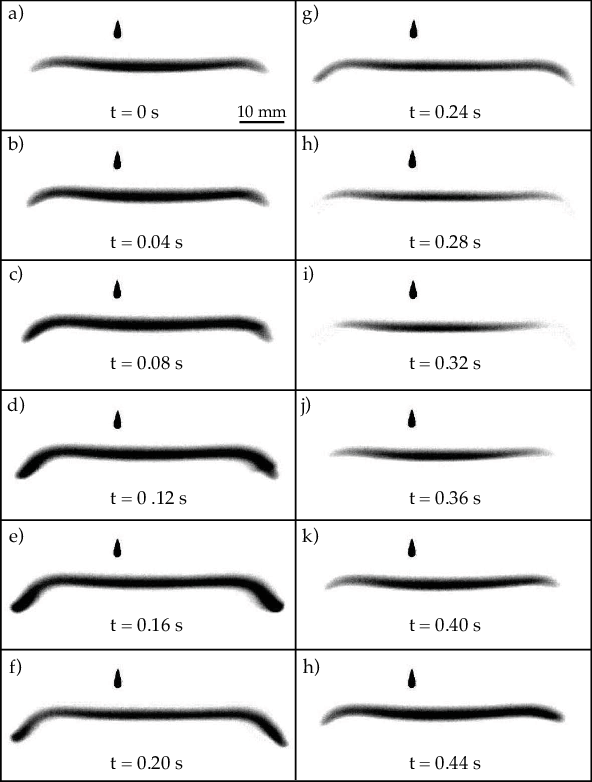}
  \end{center}
  \caption[Pulsation  in  a methane  flame  without  the inner  quartz
  cylinder present.]{Pulsation cycle of  a methane flame without inner
    quartz  cylinder.  The  corresponding  video at  1/5 the  original
    speed  is  available in  the  supplemental  material (video  \#3).
    Another   video   \#4   shows   the  reduction   of   flame   edge
    \emph{flapping} at increased bulk velocity. The dot visible at the
    top of the  images it the tip of  the mass spectrometer capillary,
    located $3$mm upstream of  the injection array. The parameters for
    this  figure,  video  \#3  and   video  \#4  are  given  in  table
    \ref{tab:ListPulsFlam}.}
  \label{fig:CH4Puls}
\end{figure}

When  the oscillatory  instability develops  without the  inner quartz
cylinder in  the combustion chamber,  the flame edges  become unstable
first and dip downward during part of the oscillation cycle as seen in
the sequence  of Fig.  \ref{fig:CH4Puls}: During the  high intensity
phase, the  flame expands towards  the windows of the  burning chamber
where the locally  lower mixture strength pushes the  flame edges down
towards the  fuel side.   The phenomenon is  less pronounced  when the
flame pulsates at  high frequency. For instance, if  the bulk velocity
is  increased from  the $13.87$  mm/s in  Fig.  \ref{fig:CH4Puls} to
$19.42$ mm/s, the flame  edges remain almost stationary throughout the
cycle, as  seen in video \#4  of the supplemental  material. A careful
comparison  of oscillating flames  with and  without the  inner quartz
cylinder showed no significant differences of frequency and luminosity
amplitude  in the  central part  of the  flame at  identical operating
parameters.   Therefore, most  results of  this section  were obtained
without the inner cylinder which  allowed the investigation of a wider
parameter range without causing rapid extinction.

% Fig. 7.16 : Pulsation without cylinder
% Serie  16.8.5.2 : 13.87 mm/s,  40% he, 2.62 hz,  11.5% CH4, phi=0.62
% (Wings)
% Serie  16.10.11.2 : 19.42 mm/s,  75% He, 5.81 hz,  11% CH4, phi=0.95
% (Low wings)
%*********(THIS VIDEO PLUS ``NO WINGS'' VIDEO IN SUPP MATERIAL)*******
% Fig. 7.13 : Sampling window
\begin{figure}
  \begin{center}
    \includegraphics[width=0.48\textwidth]{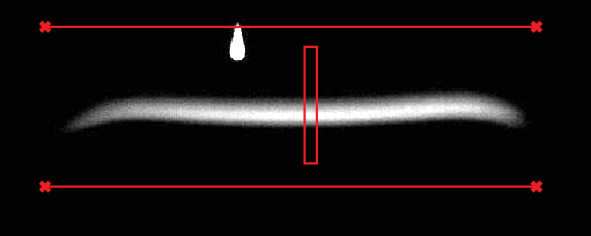}
  \end{center}
  \caption[Example   of   sampling   window   used   for   the   video
  analysis.]{Example of sampling window (rectangle) used for the video
    analysis. Horizontal lines are  reference marks located at the tip
    of the injection arrays ($20$mm chamber length).}
  \label{fig:Samplingwindow}
\end{figure}

The  videos of  the pulsating  flames  have been  analyzed to  extract
frequency and  amplitude of the  pulsations. Thereby it  is understood
that  the amplitude  of the  light emission  obtained from  the  8 bit
brightness  signal  is  only  an  indirect  measure  of  the  chemical
activity.  In  particular, zero  brightness cannot be  identified with
extinction  because of  the limited  dynamic range  of the  camera. To
avoid pixel saturation and obtain  a signal over the entire cycle, the
exposure time  and the aperture had  to be adapted from  case to case.
As a  result, the  absolute light emission  signal cannot  be compared
reliably between  different flames.  However, the image  sensor of the
camera was used with a linear sensitivity setting ($\gamma=1$) so that
the amplitude of  the emitted light intensity relative  to its average
was  considered the best  measure to  compare pulsation  amplitudes of
different flames. This relative  signal was acquired within a sampling
window as  shown in Fig. \ref{fig:Samplingwindow}.  For  each of the
10-20 pixel columns in this  window, the magnitude and position of the
intensity  maximum was  found and  between  500 and  5000 frames  were
processed for  each flame.  To  determine the pulsation  frequency and
average amplitude  of flame oscillations, the raw  intensity signal is
first filtered with a FIR  band-pass filter designed in MATLAB to pass
both the  fundamental frequency  and its first  harmonic which  can be
used  to characterize the  degree of  nonlinearity of  the oscillating
flame system.   Then the  signal is divided  into blocks of  1048 data
points and Hann-windowed providing a side lobe attenuation of $-32$db.
The  amplitude  attenuation  of  this  processing  chain  was  finally
determined  by processing a  synthetic signal  of known  amplitude and
frequency similar to the flame frequency.

\begin{figure}[ht!]
  \begin{center}
    \subfigure[]
    {
      \label{fig:PulsInt:a}
      % Serie 16.8.10.1
      \includegraphics[width=0.22\textwidth]{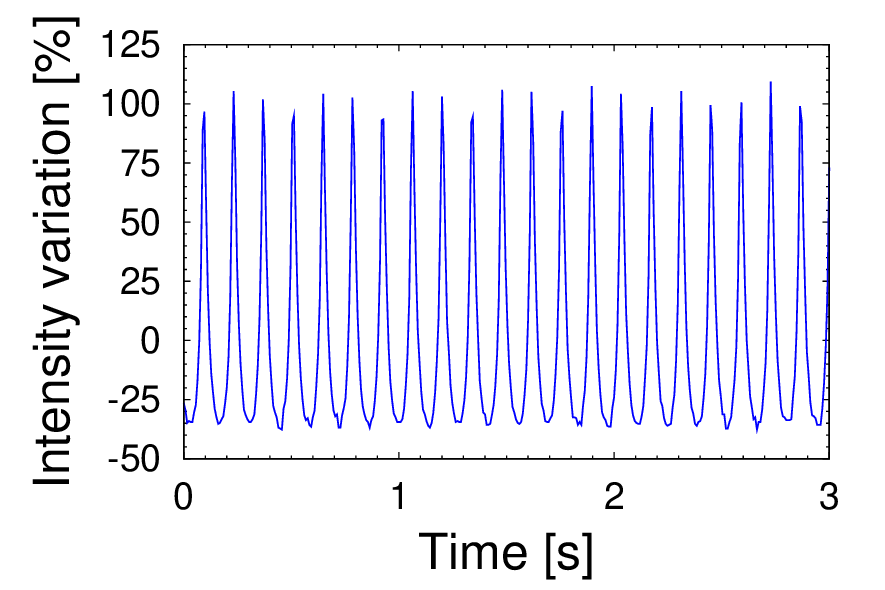}
    }    
    \subfigure[]
    {
      \label{fig:PulsInt:b}
      % In cylinder, serie 14.5.1.3
      \includegraphics[width=0.22\textwidth]{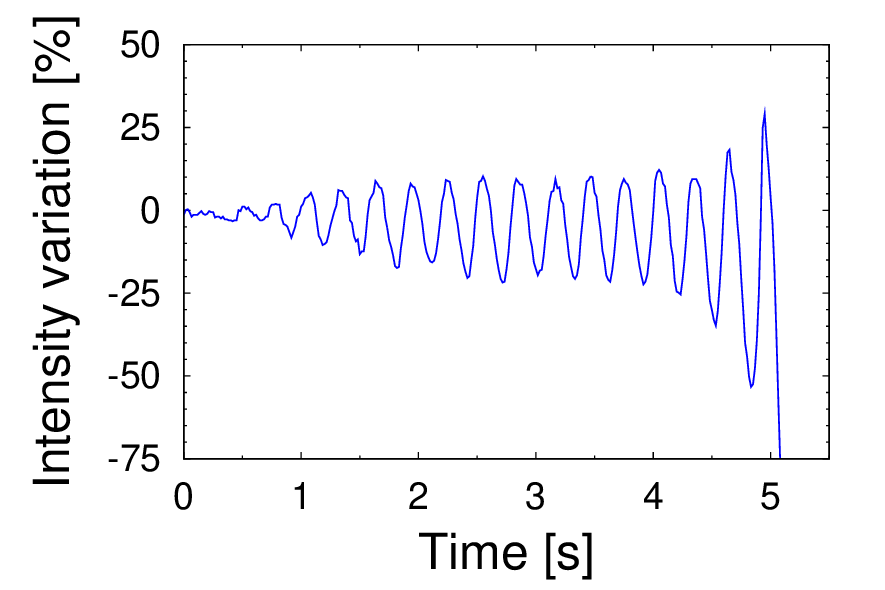}
    }
    \subfigure[]
    {
      \label{fig:PulsInt:c}
      % Serie 16.8.21.2
      \includegraphics[width=0.22\textwidth]{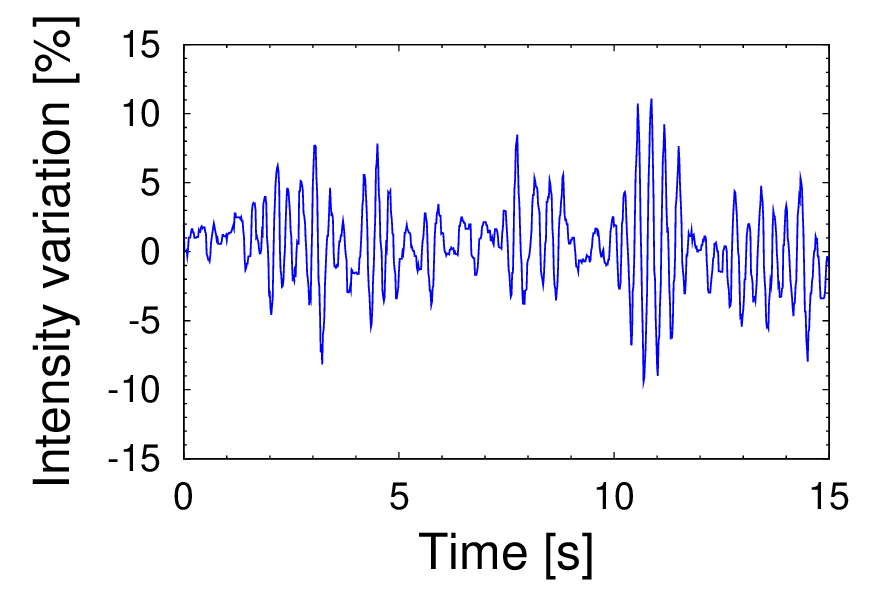}
    }
    \subfigure[]
    {
      \label{fig:PulsInt:d}
      % Serie 16.8.13.2
      \includegraphics[width=0.22\textwidth]{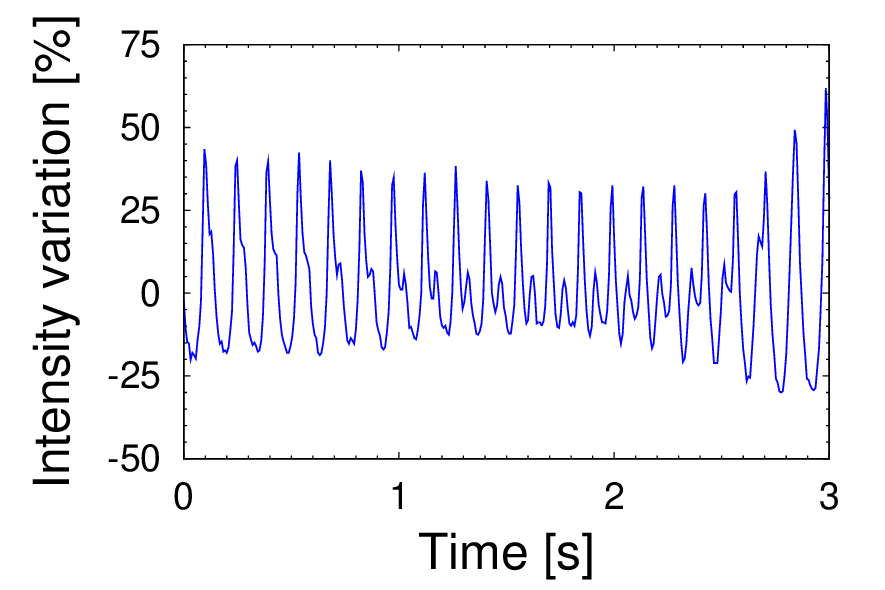}
    }
  \end{center}
  \caption[Variation  of the  flame intensity  for different  types of
  pulsations.]{  Emitted light  intensity relative  to  mean intensity
    versus  time.  a)  Stable  limit cycle;  b)  Pulsation with  rapid
    growth  in  amplitude   leading  to  extinction;  c)  Intermittent
    pulsations close to the  marginal stability; d) Pulsation with two
    competing frequencies.}
  \label{fig:PulsInt}
\end{figure}

Examples  of the  relative light  emission  signal are  shown in  Fig.
\ref{fig:PulsInt}. For the majority of flames the signal was that of a
saturated  limit cycle,  as illustrated  by  Fig. \ref{fig:PulsInt:a},
which  remained  stable for  hours,  i.e.   in practice  indefinitely.
Exceptionally, however,  different behavior  was observed: In  some of
the  flames the onset  of the  instability was  quickly followed  by a
period of  rapid growth of  the pulsation amplitude followed  by flame
extinction  as  in   Fig.  \ref{fig:PulsInt:b}.   This  behavior  is
particularly  common in  flames pulsating  at low  frequency  (1-3 Hz)
(hydrogen flames, flames with low  bulk flows) and in flames where the
inner quartz cylinder is present.  In other flames at the threshold of
instability  it  was  possible  to observe  pulsations  appearing  and
decaying  spontaneously due  to  minute variations  of conditions.  An
example  of  such  an   intermittent  pulsation  is  shown  in  Fig.
\ref{fig:PulsInt:c}.    Spatial    inhomogeneities   in   the   burner
occasionally  resulted  in  two  regions  of the  flame  pulsating  at
slightly different  frequencies, producing a modulated  signal as seen
in Fig.  \ref{fig:PulsInt:d}, or also regions pulsating  at the same
frequency  but  out  of  phase.   However, these  behaviors  were  the
exception and for the majority of  the flames the signal was that of a
saturated limit  cycle, as illustrated  by Fig. \ref{fig:PulsInt:d},
which remains stable and sometimes was observed for over one hour.

% Fig. 7.18 : Sample signals

\begin{figure}
  \begin{center}
    \subfigure[]
    {
      \label{fig:PulsIntPos:a}
      \includegraphics[width=0.4\textwidth]{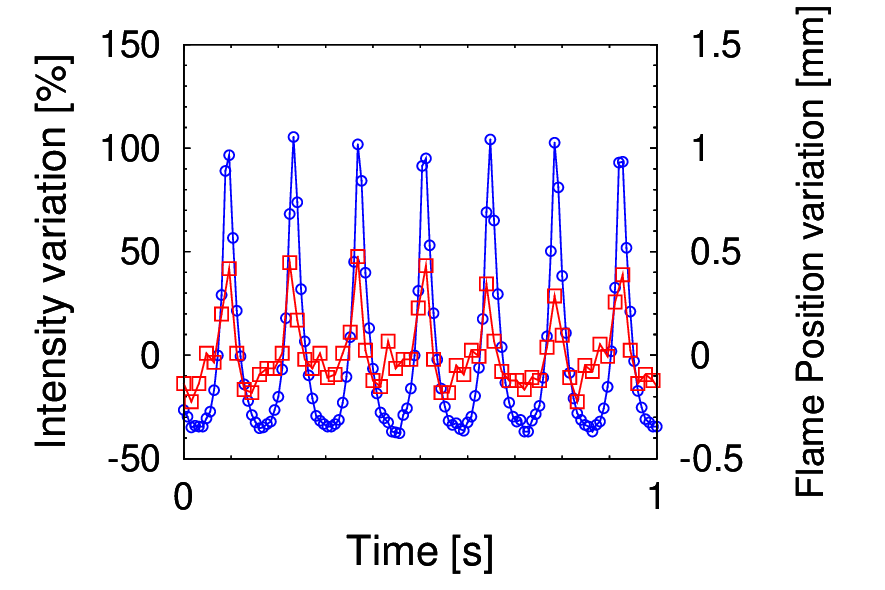}
    }
    \subfigure[]
    {
      \label{fig:PulsIntPos:b}
      \includegraphics[width=0.4\textwidth]{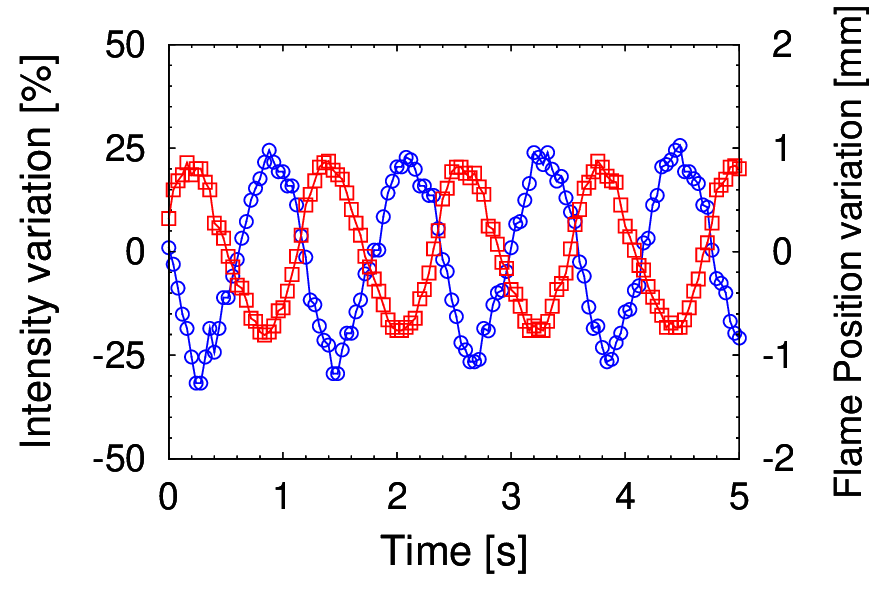}
    }
 \end{center}
 \caption[Simultaneous   variation   of   the  flame   intensity   and
 position.]{Simultaneous  traces  of  flame  intensity  ($\odot$)  and
   position ($\Box$).  a) High frequency  pulsations at $7.21$ Hz ; b)
   Low frequency pulsations at $0.86$ Hz.}
  \label{fig:PulsIntPos}
\end{figure}

% Variation of flame position
As the chemical  reaction rate oscillates, so does  the heat released.
This has an effect on the fluid mechanics of the burner and results in
a coupling  between flame intensity  and position. Thereby one  has to
distinguish  between  high  velocity  flames where  the  periodic  gas
expansion  due  to the  oscillating  heat  release  causes only  small
oscillations of  the flame  position, and flames  with slow  bulk flow
where   the   effect   is    large.    This   is   shown   in   Fig.
\ref{fig:PulsIntPos}: In  the high velocity  flame on the  left, which
pulsates at the relatively high frequency of 7.21 Hz, the amplitude of
the flame position is of the order  of $\pm 0.3$ mm and not visible to
the  naked eye.   For the  low bulk  flow on  the right  the pulsation
frequency is below 1 Hz and  the position amplitude is much larger, of
the order of $\pm 1$ mm.

% Fig. 7.19 : Pulsation of intensity and flame position
% Pos_High_16.8.3.2.dat
% Pos_Low_16.8.10.1.dat

\begin{figure}
  \begin{center}
    \includegraphics[width=0.48\textwidth]{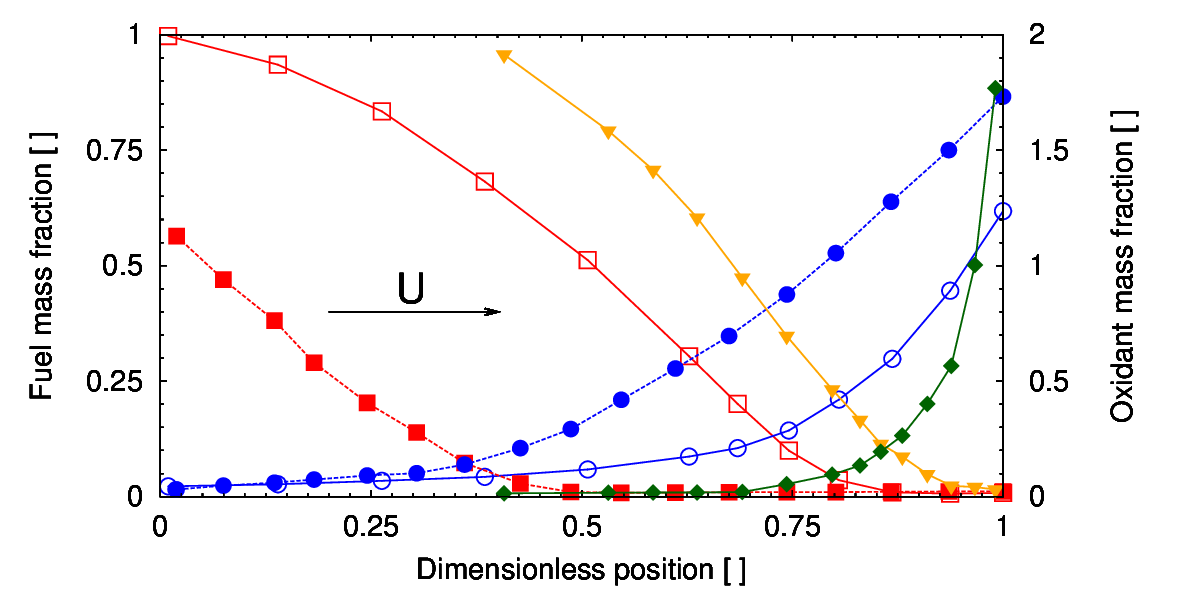}
  \end{center}
  \caption[Concentration     profiles     across     two     pulsating
  flames.]{Reactant   concentration  profiles  showing   the  reactant
    leakage   across  two  pulsating   flames  with   different  inert
    compositions.            (\textcolor{red}{$\Box$}),          fuel;
    (\textcolor{blue}{$\odot$}),  oxidant  concentrations.  Parameters
    corresponding   to   open   and   solid  symbols   as   in   table
    \ref{tab:ListPulsFlam}    \ref{fig:ConcProfPuls}(a)    and    (b),
    respectively. Also shown  are concentration profiles taken through
    a   lean   ($\phi=0.5$)  stable   planar   hydrogen  flame   sheet
    (\textcolor{orange}{$\filledtriangledown$})                   fuel;
    (\textcolor[rgb]{0,0.4,0}{$\filleddiamond$} oxidant; corresponding
    to frame b) in Fig. \ref{fig:CellularFlames}), very close to the
    onset of the first cells. The lines are experimental fits.}
\label{fig:ConcProfPuls}
\end{figure}

As opposed to cellular flames, the flame remains spatially homogeneous
during  intensity   pulsations,  at  least  in   its  center.   Hence,
representative  species concentration  profiles  can be  taken in  the
burner along  its axis.  Two such  longitudinal concentration profiles
in  flames performing  high  frequency oscillations  with small  flame
front displacements,  of the  order of the  spatial resolution  of the
mass spectrometer, are  shown in Fig. \ref{fig:ConcProfPuls}.  These
profiles  reveal an  important  reactant leakage  across the  reaction
zone, similar to the leakage through flames very close to the onset of
cellular instability,  an example  of which is  also included  in this
figure.  This means  that there  is a  significant degree  of reactant
pre-mixing in  the three  flames of Fig.  \ref{fig:ConcProfPuls} and
hence in all the unstable flames studied here, while further away from
extinction  no leakage  was observed  \cite{Robert2008}.   This means
that  critically  stable  and  unstable  flames are  at  least  partly
premixed  due to leakage.  Conversely one  can say  that leakage  is a
prerequisite for TDI in diffusion flames.

% Fig. 7.20 : Concentration profiles in pulsating flame, with leakage

% Table    for    flame     parameters    in    figure    fig:CH4Puls,
% fig:CH4Puls_NoWings, fig:PulsInt, fig:PulsIntPos, fig:ConcProfPuls.
\begin{table*}[ht!]
\begin{center}
\begin{small}
\begin{tabular}{|| l || c | c | c | c | c | c | c | c ||}
\hhline{|t:=:t:========:t|}
Figure & $U$ & \% CH$_4$ & \multicolumn{2}{c|}{Inert}& $Le_o$ & $Le_f$ &
$\phi$ & $f$ \\
\hhline{||~||~~--~~~~||}
 & [mm/s] & & [\%He] & [\%CO$_2$] & [ ] & [ ] & [ ] & [hz]\\
\hhline{|:=::========:|}
% Serie 16.8.5.2
\ref{fig:CH4Puls} and supp video \#3& $13.87$ & $11.5$ &$40$ & $60$ & $1.05$ & $0.95$ &
$0.62$ & $2.62$\\
\hhline{||-||--------||}
% Serie 16.10.11.2
Supp. Video \#4 & $19.42$ & $11.0$ &$75$ & $25$ & $1.23$ & $1.08$ &
$0.95$ & $5.81$\\
\hhline{||-||--------||}
% Serie 14.5.1.3
\ref{fig:PulsInt:a} &  $19.42$ & $11.0$ &$75$ & $25$ &  $1.23$ & $1.08$ &
$0.95$ & $5.81$\\
\hhline{||-||--------||}
 % Serie 16.8.21.2
\ref{fig:PulsInt:b} & $13.87$ & $11.0$ & $15$ & $85$ & $0.96$ & $0.88$ &
- & $1.04$\\
\hhline{||-||--------||}
% Serie 16.8.10.1
\ref{fig:PulsInt:c} & $13.87$ & $12.5$ & $90$ & $10$ & $1.41$ & $1.16$ &
$1.31$ & $7.21$\\
\hhline{||-||--------||}
% Serie 16.8.13.2
\ref{fig:PulsInt:d} & $13.87$ & $12.5$ & $85$ & $15$ & $1.33$ & $1.13$ &
$1.18$ & $6.97$\\
\hhline{||-||--------||}
% Serie 16.8.10.1
\ref{fig:PulsIntPos:a} & $13.87$ & $12.5$ & $90$ & $10$ & $1.41$ & $1.16$ &
$1.31$ & $7.21$\\
\hhline{||-||--------||}
% Serie 16.8.3.2
\ref{fig:PulsIntPos:b} & $13.87$ & $12.0$ & $20$ & $80$ & $0.95$ & $0.86$ &
$0.89$ & $0.86$\\
\hhline{||-||--------||}
% Serie 16.9.1
\ref{fig:ConcProfPuls}(a) & $13.87$ & $12.0$ & $50$ & $50$ & $1.09$ & $0.99$ &
$0.81$ & $3.67$\\
\hhline{||-||--------||}
% Serie 16.9.2
\ref{fig:ConcProfPuls}(b) & $13.87$ & $12.0$ & $0$ & $100$ & $0.91$ & $0.82$ &
$0.57$ & $0.73$\\
\hhline{|b:=:b:========:b|}
\end{tabular}
\end{small}
\end{center}
\caption[Burner  parameters  for   the  flames  presented  in  section
\ref{sec:ObservationsPuls}.]{Parameters    for   the figures of
  section \ref{sec:Pulsations}. The symbol [-] signifies that
  data are not available.}
\label{tab:ListPulsFlam}
\end{table*}
\normalsize

\subsection{Scaling of pulsation frequency }
\label{sec:PulsScaling}

The  linear stability analysis  of thermal-diffusive  instabilities in
the  idealized one  dimensional  configuration\cite{Kukuck2001} yields
the pulsation frequency

\begin{equation}
f \sim \frac{U^2}{2 \pi D_{th}} \omega_I^\ast
\label{eq:PulsFreq}
\end{equation}

where $\omega_I$ is the  imaginary part of the non-dimensional complex
frequency,  i.e.   the  oscillation  frequency of  the  most  unstable
disturbance  at  marginal  stability.    Little  is  known  about  the
dependence of $\omega^*_I$ on $Da$, $\phi$, $Le_f$, $Le_o$ and $\theta$,
where  $\theta$  is  the  activation  energy  parameter  or  Zeldovich
number. In  the literature \cite{Kukuck2001, Metzener2006}  only a few
values of  $\omega_I$ are  reported for parameter  sets which  are far
from those of the present  experiment.  As $f D_{th}/U^2$ was found to
be far  from constant, $\omega_I$, contrary to  $\sigma^*$ in equation
(\ref{eq:CellSize}), must vary  significantly over the parameter range
of  the  present  experiments.   This variation  is  now  investigated
experimentally.   It turns  out  that $f  D_{th}/U^2$ correlates  most
consistently with the Damk\"ohler number and the pulsation amplitude.

The  determination   of  the  Damk\"ohler   number  from  experimental
parameters is non-trivial, as a large number of parameters is involved
in its definition \cite{Cheatham2000}

\begin{equation}
\mathcal{D}_a  =  \frac{\lambda}{\rho_a   c_p  U^2}  \left(  \frac{R^0
    T_a}{E} \right)^3  \frac{\nu_x c_p \bar{W}}{q  R^0W_f} \mathcal{B}
Y_{f,0} p_0 \exp(-E/RT_a)
\label{eq:Dam2}
\end{equation}

In  this  definition, $\rho$,  $c_p$,  $\lambda$,  $U$  and $T_a$  are
readily available from the  experiment. The main difficulty resides in
finding suitable \emph{effective} values for the activation energy $E$
and  the Arrhenius  pre-exponential factor  $\mathcal{B}$ representing
the  global  reaction.   One  can  find in  the  literature  data  for
elementary  steps  of  many  reactions  that can  be  complemented  by
estimates obtained analytically  from statistical mechanics.  However,
when modeling  a whole  complex mechanism by  a global  reaction these
values  need to be  determined empirically.  Fortunately $\mathcal{B}$
exhibits  only a weak  temperature dependence  and therefore  does not
significantly   contribute  to  the   variation  of   the  Damk\"ohler
number.  The terms  expected to  have a  significant influence  on its
variation  between the  flames studied  here are  $\lambda/(\rho_a c_p
U^2)=D_{th}/U^2$,  $(R^0 T_a/E)^3$  and  most importantly  $exp(-E/R^0
T_a)$. Lacking  suitable data, all the other  parameters including $E$
are assumed constant.  The Damk\"ohler number used in the following is
therefore a relative one, calculated with respect to a reference flame
arbitrarily selected  from the  available data for  which $Da$  is set
equal to unity.

\begin{figure}[ht!]
  \begin{center}
    \subfigure[]
    {
      \label{fig:ScalingDa:a}
      \includegraphics[width=0.45\textwidth]{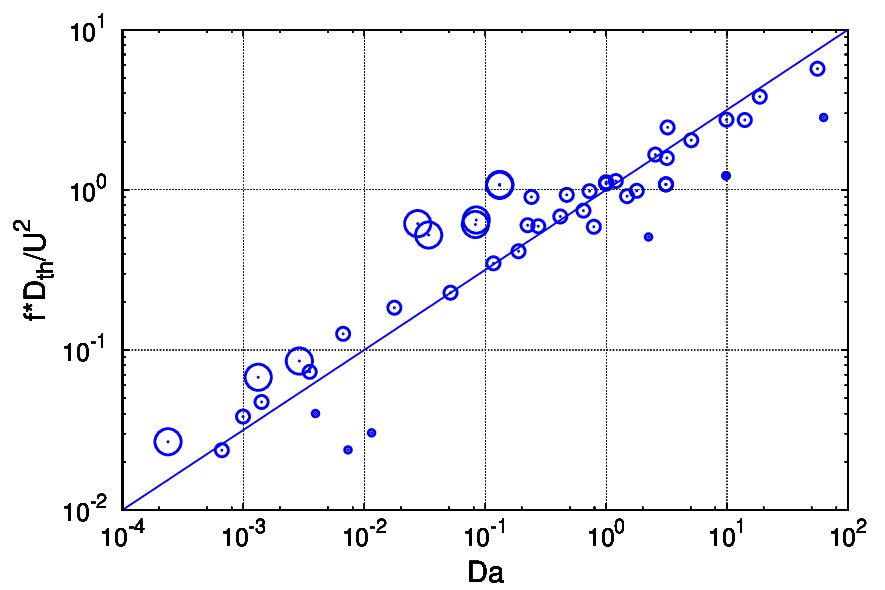}
    }
    \subfigure[]
    {
      \label{fig:ScalingDa:b}
      \includegraphics[width=0.45\textwidth]{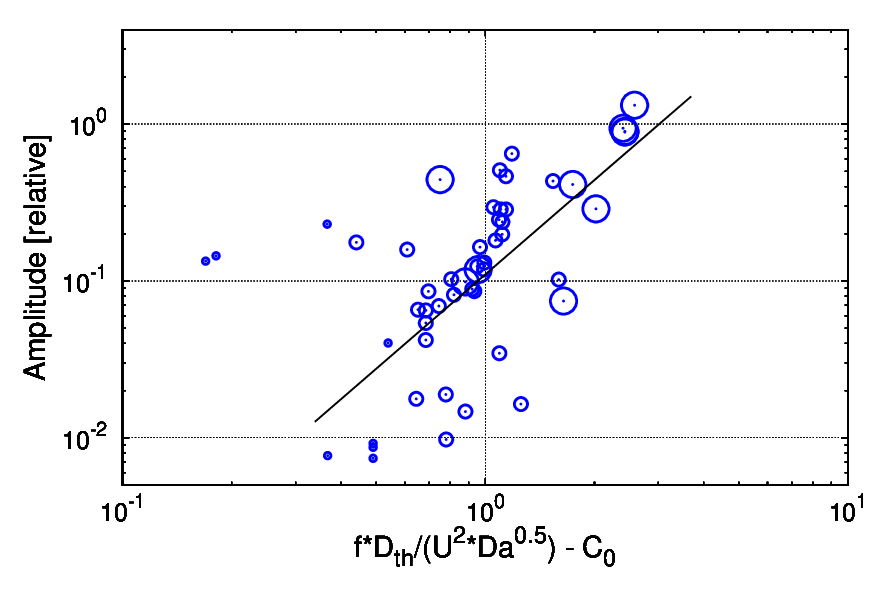}
    }
  \end{center}
  \caption[Scaling  of the  pulsation frequency  with  the Damk\"ohler
  number.]{a)  Non-dimensional  pulsation  frequency  versus  relative
    Damk\"ohler number. Open and  solid symbols correspond to cases at
    constant     inert    composition    and     constant    velocity,
    respectively.  Symbol  size   is  proportional  to  the  amplitude
    $\epsilon$.   ----,  $f  D_{th}/U^2   =  Da^{1/2}$.   b)  Equation
    (\ref{eq:NLStab})  and data  of Fig.  \ref{fig:ScalingDa:a} with
    $C_0 = 0.05$.}
  \label{fig:ScalingDa}
\end{figure}

The measured dimensionless frequency $f*D_{th}/U^2$ is shown in Fig.
\ref{fig:ScalingDa:a}  as  a  function  of this  relative  Damk\"ohler
number. At  the same time  the amplitude $\epsilon$ of  the pulsation,
defined as  the difference  between the peak  and average  flame light
emission, is visualized on this graph by the size of the symbols which
is  proportional   to  $\epsilon$.   A  close   inspection  of  Fig.
\ref{fig:ScalingDa:a}  reveals that,  for flames  \emph{oscillating at
  comparable amplitudes}, $f*D_{th}/U^2$ is reasonably proportional to
$Da^{1/2}$. However, it also appears that the proportionality constant
increases with  amplitude. This is  not entirely unexpected  since the
measured  frequency is  that of  saturated  limit cycles  and not  the
frequency  $\omega_I$  of   infinitesimal  amplitude  oscillations  in
equation  (\ref{eq:PulsFreq}).  The limit-cycle  oscillations observed
here are believed to be the result of a supercritical Hopf bifurcation
\cite{DiBenedetto2002}.   If so,  the weakly  non-linear Stuart-Landau
theory \cite{Stuart1971} applies close to the bifurcation. It predicts
that both the  square of the oscillation amplitude  and the difference
between  non-linear  and  linear  frequency are  proportional  to  the
bifurcation parameter,  i.e. the distance from  the bifurcation. Hence
we  expect  the  following   relation  between  observed  limit  cycle
frequency, $Da$ and the pulsation amplitude $\epsilon$:

\begin{equation}
\frac{f D_{th}}{U^2 \sqrt{Da}} = C_0 + a \epsilon^2
\label{eq:NLStab}
\end{equation}

The  constant   $C_0$  is  the  dimensionless   frequency  divided  by
$\sqrt{Da}$ at zero amplitude,  corresponding to the marginally stable
state.   Its value  could  be estimated  directly  from the  available
experimental data  at very low  amplitude as $C_0 \approx  0.05$.  The
relation (\ref{eq:NLStab})  is plotted together  with the experimental
data in Fig. \ref{fig:ScalingDa:b}. The value of the coefficient $a$
was then determined  by least squares. As is  evident from the figure,
the  optimal  value $a=100$  did  only result  in  an  $R^2$ value  of
approximately $0.6$.   It should be  noted however that  the amplitude
used here  is only  an indirect measure  of the amplitude  of reaction
rate  or temperature  oscillations which  is in  addition  affected by
spatial averaging inherent to the acquisition technique used here (see
section \ref{sec:Pulsations}).

% Fig. 7.24 : Scaling and amplitude

\section{Conclusions}
\label{sec:Conclusions}

% The experimental results on instabilities
The   main   contribution   of   this  study   is   the   quantitative
characterization  of  thermal-diffusive  instabilities  of  unstrained
diffusion flames  close to  extinction.  The use  of a mixture  of two
inert gases (He$_2$ and CO$_2$) to dilute the reactants has allowed to
map the type of instabilities  arising close to extinction over a wide
range  of Lewis  numbers for  both  hydrogen and  methane flames.   In
hydrogen  flames,  a cellular  structure  was  observed  at low  Lewis
numbers  produced with  a  dilution of  the  reactants principally  by
CO$_2$.  The characteristic size  of the cellular pattern was observed
to  increase  with  Lewis  number,  approximately  doubling  when  the
diluting inert  CO$_2$ was  completely replaced by  He. The  cell size
scaled  close to  linearly with  the diffusion  length $l_D$,  in good
agreement with theoretical predictions.

The  transition from  cellular  to intensity  pulsations occurred  for
hydrogen flames when $80$ \%  of the dilution mixture was helium.  The
weak light  emission of these  H$_2$ pulsating flames  prevented their
detailed investigation  and methane flames were used  instead to study
flame  pulsations for  the  first time  in  an essentially  unstrained
setting.   Pulsation frequencies  in  the range  of  $0.6-11$ Hz  were
observed using  the same two inerts for  dilution. Lacking theoretical
guidance, the scaling of the pulsation frequency with flame parameters
was  developed  from experiment.   While  the  proportionality of  the
non-dimensional  saturated frequency  $f D_{th}/U^2$  with  the square
root of the Damk\"ohler  number, $Da^{1/2}$, could be established with
some  confidence,  the  dependence  of frequency  on  the  oscillation
amplitude  according to  weakly non-linear  theory must  be considered
more tentative.

% Pertinence of the experimental results produced
The  present  experimental  data   provide  for  the  first  time  the
possibility  of  validating  theoretical  models. However,  to  really
arrive at quantitative comparisons,  theoretical analyses will have to
be  adapted to  take into  account  the practical  limitations of  the
experiment  such as  the  coupling between  Lewis  numbers, bulk  flow
velocity and mixture  strength as well as the  finite amplitude of the
observed instabilities.

\section*{Acknowledgments}
\label{sec:Ack}

The  authors wishes  to  express their  gratitude  to Professor  Moshe
Matalon for his valuable  comments and insights regarding the material
discussed  in this  paper.  Additionally,  financial support  for this
research was received from the Swiss National Science Foundation under
Grant  20020-108074 and from  the Swedish  Research Council  (VR grant
2009-6159).

\section*{Nomenclature}
\label{sec:Nom}

\hspace{-0.5cm}
\begin{center}
  \begin{supertabular}[htbp]{l l p{4cm}}
    $\mathcal{B}$ & [m$^3$/mol*s] & Arrhenius pre-exponential factor\\
    $C_p$ & [J/kg*K] & Specific heat at constant pressure\\
    $D_a$ & [ ] & Damk\"ohler number\\
    $D_i$ & [m$^2$/s] & Diffusivity of species $i$\\
    $D_{th}=\lambda/(\rho C_p)$ & [m$^2$/s] & Thermal diffusivity\\
    $E$ & [J/mol] & Activation energy\\
    $f$ & [1/s] & Pulsation frequency\\
    $Le_i=\kappa/D_i$ & [ ]& Lewis number of species $i$\\
    $p_0$ & [Pa] & Ambient pressure\\
    $q$ & [J/kg] & Heat released per unit mass of fuel\\
    $Q$ & [J/mol] & Total heat released\\
    $R$ & [J/kg*mol] & Ideal gas constant\\
    $T_a$ & [K] & Adiabatic flame temperature\\
    $U$ & [m/s] & Bulk flow velocity\\
    $\bar W$ & [kg/mol] & Mean molecular weight\\
    $W_i$ & [kg/mol] & Molecular weight of specie $i$\\
    $Y_f$ & [ ] & Fuel mass fraction\\
    $Y_o$ & [ ] & Oxidizer mass fraction\\
    $x$ & [m] & Coordinate along burner length\\
    $\epsilon$ & [variable] & Pulsation amplitude\\
    $\lambda$ & [W/m*K] & Thermal conductivity\\
    $\nu_i$ & [ ]& Stoichiometric coefficient of species $i$\\
    $\phi$ & [ ]& Equivalence ratio (mixture strength)\\
    $\rho$ & [kg/m$^3$] & Density \\
    $\sigma^\ast$ & [ ] & Real part of critical growth rate in the
    marginal state\\
    $\omega$ & [mol/s] & Chemical reaction rate\\
    $\omega_I^\ast$ & [ ] & Imaginary part of critical growth rate in the
    marginal state\\
  \end{supertabular}
\end{center}

% Normal Bibliography
%--------------------

% BibTex Bibliography
%--------------------
%\bibliographystyle{elsarticle-num}
%\bibliography{Bib2010}

\end{document}